# From Nano-Communications to Body Area Networks: A Perspective on Truly Personal Communications

**PAWEL KULAKOWSKI[1], KENAN TURBIC[2], (Member, IEEE), and LUIS M. CORREIA[2], (Senior Member, IEEE)**
[1]AGH University of Science and Technology, Krakow, Poland
[2]INESC-ID/Instituto Superior Técnico, University of Lisbon, 1000-029 Lisbon, Portugal

Corresponding author: Pawel Kulakowski (e-mail: kulakowski@kt.agh.edu.pl).

This work was developed within the framework of the COST Action CA15104, IRACON. It was also partially supported by the Polish Ministry of Science and Higher Education with the subvention funds of the Faculty of Computer Science, Electronics and Telecommunications of AGH University of Science and Technology.

**ABSTRACT** This paper presents an overview of future truly personal communications, ranging from networking inside the human body to the exchange of data with external wireless devices in the surrounding environment. At the nano- and micro-scales, communications can be realized with the aid of molecular mechanisms, Förster resonance energy transfer phenomenon, electromagnetic or ultrasound waves. At a larger scale, in the domain of Body Area Networks, a wide range of communication mechanisms is available, including smart-textiles, inductive- and body-couplings, ultrasounds, optical and wireless radio transmissions, a number of mature technologies existing already. The main goal of this paper is to identify the potential mechanisms that can be exploited to provide interfaces in between nano- and micro-scale systems and Body Area Networks. These interfaces have to bridge the existing gap between the two worlds, in order to allow for truly personal communication systems to become a reality. The extraordinary applications of such systems are also discussed, as they are strong drivers of the research in this area.

## I. INTRODUCTION

The several successive generations of mobile cellular and wireless communications have been aiming at a single goal: to provide connectivity to people at their own pleasure. It started with the famous "anytime, anywhere" motto in the 1st Generation of voice-only mobile cellular communications, and since then it has evolved to the provision of data and multimedia everywhere and with a decreasing delay. In addition to satisfying the anticipated data rate requirements, the incoming 5th Generation aims at two other key goals: to reduce transmission latency and to substantially increase network capacity. These two goals are not improving the direct user's experience, but rather enabling machine-based applications and services. So, somehow, the further evolution of mobile and wireless communications has to eventually address the truly personal dimension of communications, i.e., the exchange of information within, around, and outside the body.

Body area networks (BANs), accommodating such communication scenarios, have been gaining considerable attention recently. However, for truly personal communication systems, one needs to encompass nano-networks [1], to allow for the exchange of information among devices inside the human body, by exploiting mechanisms at the cellular and molecular levels. At the nano-scale, communications are performed by using molecules or molecular structures with specific properties, such as photo-active fluorophores or channelrhodopsins, antibodies (proteins), moving bacteria or waves of ions at different scales, as shown in Fig. 1. Of course, this ultrawide range of communication mechanisms introduces a number of challenges, including those related to power supply, propagation delay, and throughput, among many others. These challenges need to be addressed from the technical aspect, for example, design of antennas, electronics, interfaces, materials, software, etc., but also from the socio-psychological one, as the deployment of these systems will highly depend on acceptance by society.

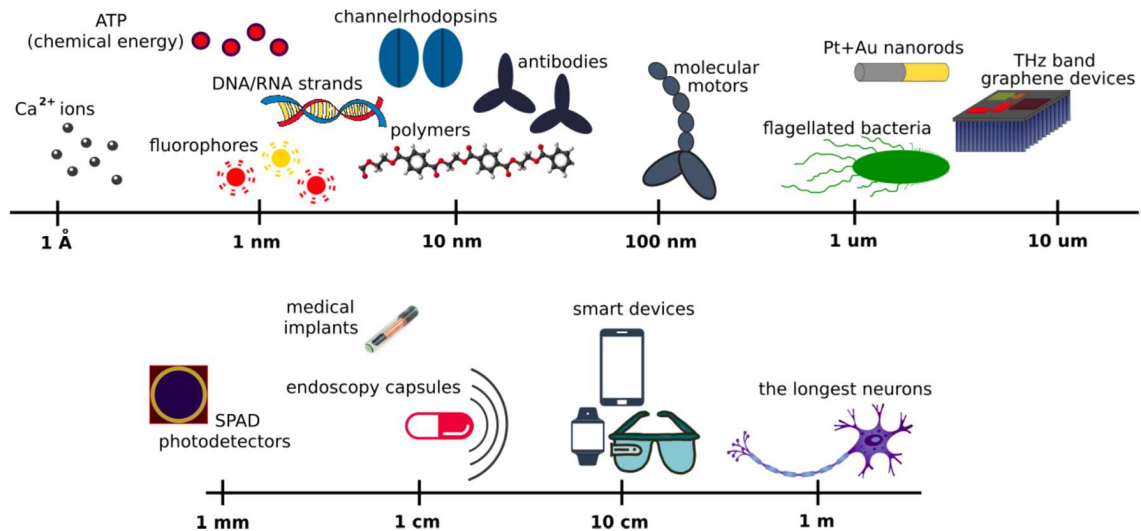

**FIGURE 1.** Scale of communication mechanisms and devices.

On the other hand, since these systems will be an integral part of the body, they may truly change people's lives. Imagine if one's thoughts could be picked up and sent to another person, without the need for a verbal explanation, or if one's body could detect cancerous tissues, heart and brain function anomalies, and warn physicians well in advance of a heart attack, seizure, or other complications. While these examples stretch one's imagination, the potential of these personal systems is extraordinary.

This paper addresses these truly personal communication systems, with the focus on the aspects of integration of nano-communication systems into a large-scale communication network. A brief overview of nano- and BAN communication technologies is provided, and the potential mechanisms that can be used to establish interfaces in between the two are discussed. The latter is the main contribution of this paper, as interfaces are addressed by taking a perspective different from the typical one, by underlining their crucial role in making these systems work. Thus, beyond the reviewing of techniques in nano-networks and BANs, the paper brings together insights into physics, biotechnology and optogenetics, explaining new mechanisms that can be exploited for creating interfaces between the networks at the different scales.

The rest of this paper is organized as follows. Potential applications are discussed in Section II, and the components of the foreseen personal network are addressed in the following sections, in an increasing scale. Nano- and micro-scale communications inside the body are discussed by reviewing the most promising mechanisms in Section III, followed by an overview of the main communication scenarios and enabling technologies in current BANs in Section IV. The potential interfaces between networks of different scales are discussed in Section V, finishing with the main conclusions in Section VI.

## II. APPLICATIONS

Establishing communications between nano-, micro- and macro-networks, when possible, opens a vast area of new applications, stretching beyond the ones envisioned for nano-networks alone. The pivotal change is the fact that nano-networks may be contacted and controlled from the macro-scale world during the whole time of their operation.

The largest group of potential applications is related to medical diagnostics and surgery. One may imagine a diagnostic system composed of nano-particles deployed in the human tissue or the vascular system [2], [3]. Data gathered by the system can be transmitted outside the body to a medical doctor, through BAN connections, for real time analysis. Similarly, nano-machines could be used to perform remote surgery, with the surgeons having access to the collected data and steering their actions. One can easily envisage other applications in healthcare, encompassing patient monitoring, localized drug delivery, aging care, etc.

Data storage is another area that can greatly benefit from nano- and molecular-networks, considering the exponentially increasing amount of data mankind is producing. Deoxyribonucleic and ribonucleic acid (DNA and RNA) strands, which keep the human genome in nucleotide chains, are extremely effective in data storage. In theory, each nucleotide can keep 2 bits of information, as there are 4 different nucleotides both in DNA and RNA strands. It was recently shown that information can be artificially stored and retrieved in DNA very efficiently, keeping on average 1.57 bits per nucleotide, which corresponds to an impressive density of 215 petabytes per gram of DNA [4].

These systems could also comprise smart uniforms for fire fighters, police, and soldiers in battlefields, with the embedded sensors capable of measuring the body's vital signals or detect bullet wounds, hence, being essential for the safety of their users. Nano-sensors operating inside one's body could allow for quick diagnostics, localization and precise classification of

TABLE 1. Estimated parameters for nano- and micro-communication mechanisms.

| Mechanism | Range | Throughput | Signal speed | Technology maturity |
|---|---|---|---|---|
| FRET | 15 nm | 25 Mbit/s | 1 m/s | laboratory experiments |
| Calcium wave propagation | 300 μm[(1)] | 1 bit/s | 30 μm/s | laboratory experiments |
| Polymers | a few μm | a few kbit/s | below 1 mm/s | theoretical analysis |
| Molecular motors | up to 50 μm | 1 cargo/motor[(2)] | 1 μm/s | laboratory experiments |
| Bacteria | about 1 mm | up to 1 Mbit/bacterium | up to 20 μm/s | laboratory experiments |
| Nano-rods | 10 mm | 128 kbit/nanorod | up to 20 μm/s | laboratory experiments |
| Neurons | 1 m | 1 kbit/s | up to 120 m/s | laboratory experiments |
| EM-based micro-devices | 5 mm | 500 kbit/s | $1.5 \times 10^8$ m/s[(3)] | theoretical analysis |

(1) assuming a favorable alignment of cells where calcium ions propagate, usually the range is smaller.
(2) cargoes might be liposomes or vesicles containing DNA strands, quantum dots or micro-fabricated silicon.
(3) for human blood; it might differ in other tissues depending on the tissue refractive index; in graphene, it is 100 times smaller.

the tissue damage in case of injury, allowing for a quick reaction of a medical team, e.g., rescuing a soldier from the battlegrounds, and ultimately making the difference between life and death.

Many applications can be envisaged in sports as well, especially for training of high-performance athletes and monitoring of the fitness-related activities, including the measurement of different physiological parameters, e.g., heart rate, energy consumption, fat percentage, body water content, etc. The measurement and display of real-time information and/or the control of follow-up reports may lead sports to a new level of professional well-being and safety.

Backed by the power of business and commerce, the entertainment industry is one of the potential technology drivers. On the one hand, requirements may not be as strict as in the previous areas (a person's life is not at stake), but on the other hand, the required scale may represent a great challenge.

Some more futuristic applications include people "thinking" directly to a network. While some direct brain-computer interaction has already been achieved previously [5], [6], nano-communications could be the key factor for a real breakthrough progress. Assuming that human nerve cells can be connected via a nano-network to an external BAN, and then to other networks, a person could send thoughts and emotions directly through the network, without the need to type or vocalize them. This would represent a real integration of a person with a network, and a merge of the Internet of Things with the Internet of Nano-Things [7], which could represent the embodiment of truly personal communications, and the direction for the development of future wireless networks.

Finally, the enhancement of advanced human-computer interfaces will enable people to have access to a wide range of truly personal information. These systems will revolutionize healthcare as one knows it today, by advancing diagnostics and disease prevention far beyond today's possibilities, but also fundamentally change many other areas.

## III. NANO-COMMUNICATION MECHANISMS

There are a few distinct approaches to communications at the nano-scale, the main ones being:
1. *FRET-based*, where communications occur at distances of nanometers.
2. *Molecular*, which is a group of distinct phenomena at the scale of nano- to micrometers.
3. *Electromagnetic (EM)*, where devices can communicate at the millimeter scale.
4. *Ultrasonic photoacoustic*, over distances of the order of a hundred micrometers.

The estimated parameters for the communication mechanisms discussed in this section are summarized in Table 1.

The first approach is based on the Förster resonance energy transfer (FRET) phenomenon, which allows for energy to pass between molecules in a non-radiative manner. The unconventional nano-transmitter and receiver are in this case two neighboring molecules, rather than artificial devices. The transmitting molecule, excited to a high energy state by an external radiation or a chemical reaction, passes its energy to the receiving one. This transfer happens quite fast, i.e., within nanoseconds, and at nanometer distances, i.e., between molecules at most 20 nm apart. In terms of communication, FRET can be used for a binary "on-off" modulation, where bit '1' is realized by a FRET transfer, while '0' corresponds to no transfer [8]. One should note that FRET occurs only between spectrally matched molecules, i.e., when the nano-transmitter's emission spectrum overlaps the nano-receiver's absorption one. However, if needed, the spectral gap between the molecules can be filled in by using phonon-assisted energy transfer [9].

In addition to a very limited range, FRET is not particularly reliable. Instead of sending a signal to the receiver, the transmitting molecule may lose its energy by emitting a photon. This problem can be overcome by employing multiple-input multiple-output FRET, with multiple molecules at both communication sides, Fig. 2, thereby improving communication reliability by diversity, similarly to conventional radio communications [10]. One should also mention that signal routing in FRET-based networks is feasible, if molecules with specific properties are used, such as photo-switched fluorophores, quenchers or proteins of changeable shape [11].

The second approach exploits biological mechanisms, operating at scales ranging from nano- to micrometers [12]. Communications in cells of living organisms occur in various ways, with information carriers again not being EM waves, but groups of molecules, hence, the name molecular communications [13], [14], the great advantage being their bio-compatibility, as these mechanisms naturally occur in biological organisms.

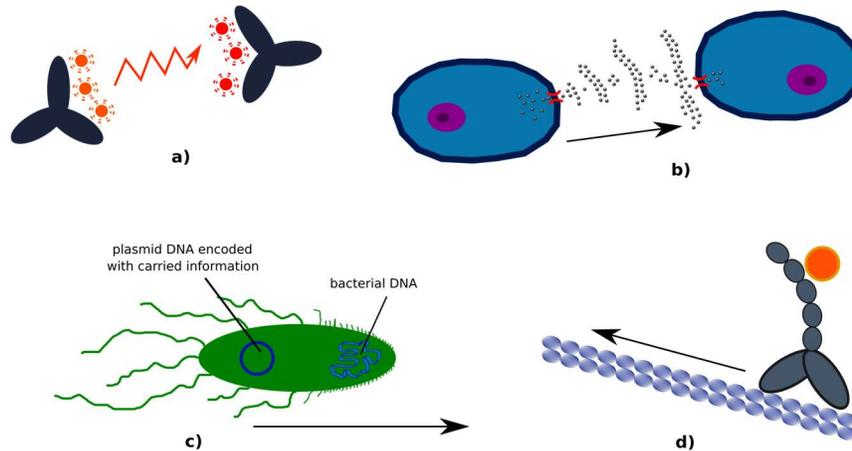

**FIGURE 2.** Nano-communication mechanisms. (a) FRET signal transfer among fluorophores mounted on antibodies.(b) Propagation of $Ca^{2+}$ ion waves among cells.(c) Bacterium moving with the force of its flagella.(d) Molecular motor moving along a micro-tubule carrying RNA data in a vesicle.

One of the most widely considered communication mechanisms is molecular diffusion. The commonly analyzed information carrier is by waves of propagating calcium ions [15], [16], information being coded as the concentration of calcium ions emitted by a cell. The ions themselves are small particles, i.e., 100 to 200 pm in size, that can propagate over distances up to 300 µm, assuming a favorable alignment of living cells for diffusion [17], [18]. Some larger particles, as polymers, can be broadcasted by diffusion as well, carrying information coded in their modified structure. For example, bits '0' and '1' can be coded by modifying polymers with either hydrogen or fluorine atoms [19]. Since the diffusion process is slow, with the average velocity of calcium ions not exceeding 30 µm/s, some diffusion acceleration methods, such as flow assisted propagation [20], are also considered.

Other molecular communication approaches include those based on molecular motors, which are about 100 nm long structures, such as kinesins or dyneins, carrying information on an RNA strand or a sequence of peptides in a vesicle [21]. These motors ensemble wired-like communications, as they move along protein tracks called micro-tubules, thus, being much slower (about 1 µm/s of propagation speed), but far more reliable than the other approaches of its class.

The size of the largest molecular carriers is of the order of a micrometer, and active transportation techniques may be used (in contrast with passive diffusion). Bits can be encoded in a DNA strand of a plasmid located inside a bacterium or attached to a catalytic nano-motor. The bacterium, such as Escherichia coli, then travels using the force of its flagella. The catalytic nano-motors, i.e., usually gold and platinum nano-rods, can exploit chemical energy from the environment by participating in chemical reactions, e.g., catalyzing the formation of oxygen [22].

As a communication medium for nano- and micro-scale, neurons, i.e., nerve cells, should be mentioned. The size of these cells varies strongly from micrometers to even 1 or 2 m (the longest ones, e.g., from a toe to the brain). The anatomy of neurons is a well-studied topic. They receive signals when molecules called neurotransmitters reach specific receptors located on a neuron membrane, which triggers the opening of ion channels on the membrane and causes some ions flow in and out of the neuron changing its inner electrical potential. The resulting electrical impulse (called action potential) travels inside the neuron till its end (called synapse), where it initiates releasing small vesicles with neurotransmitters that, via diffusion, can reach and activate another neuron or a motor cell.

Impulses in neurons travel very fast, comparing with other molecular mechanisms, reaching 120 m/s if the neuron is covered with myelin, which is a specific insulating material [23]. This makes neurons the fastest propagation medium among the considered molecular mechanisms, however, as in the case of molecular motors, signals propagate in one direction only. The achievable throughput is limited by the refractory period, which is the time after the action potential propagation when the neuron does not respond to any stimulation. The refractory period is at least 1 ms, which means that the upper bound for throughput in a single neuron is about 1 kbit/s, assuming a temporal coding of 1 bit per neural impulse (1: an impulse, 0: no impulse [24]).

The third approach is based on the idea of miniaturization of the existing EM communication techniques. The attainable scale of devices is about 10 µm, which is larger than the one in molecular networks. Therefore, the tiny size of micro-scale device restricts the communication frequency to the THz band. The use of lower and more favorable frequencies in between 0.1 and 10 THz is enabled by graphene-based antennas capable of efficiently radiating EM waves in this band [25], [26], [27], with the resonant frequencies being two orders of magnitude lower than those for the metallic antennas of the same size. On the surface of these materials, EM waves propagate as surface plasmon polaritons [28], and under specific conditions, their velocity can be even 100 times lower than in vacuum. In such circumstances, the wavelengths of THz waves are in order of 1 μm.

Communications at the THz band are associated with many challenges, the high propagation losses inside body tissues limiting the range to only a few millimeters. In addition to a high path loss, due to the spreading wave front, the rich water content of the blood and body tissues yields extremely high frequency-selective molecule absorption losses [29], resulting in path loss exceeding 120 dB for distances of a few millimeters [30]. The resonance of molecules excited by THz waves results in the conversion of EM energy into kinetic one, being absorbed by the molecule and lost from the communication perspective. On the other hand, the imposed variations of molecules result in EM radiation in the same frequency band, manifesting itself as non-white noise with power magnitude depending on the communication distance and on the number of molecules in between transmitting and receiving antennas [29]. Moreover, as molecular absorption is conditioned by the presence of a signal in the medium, the corresponding noise results in pulse spreading, and sufficient separation between successive pulses has to be ensured to prevent interference, Time Spread On-Off Keying (TS-OOK) having been proposed as a suitable coding scheme [31].

By exploiting the same mechanisms, the emerging plasmonic antennas also enable optical transmission in infra-red and visible light spectrum [32, 33], where plasmonic nano-lasers [34] and single-photon detectors [35] can serve as transmitter and receiver, respectively. As the EM waves in the THz band, optical nano-communications inside the human body are faced with challenges, but have the advantage of the molecular absorption losses in water (i.e., the majority of blood content) being minimal within the optical window. In blood vessels, the optical EM waves propagate through lossy homogeneous blood plasma and interact with different cells, particularly with the red blood cells (erythrocytes), being the most abundant ones [36]. Propagation is dominated by refraction through the cells in between transmitter and receiver, and by reflection from the surrounding ones [37], resulting in the multipath effect. However, red blood cells are reported to focusing the light propagating through them [37, 38], thus reducing the exponential path loss in favor of communications. The observed dependence of the focusing properties on the shape and orientation of cells can be potentially used to detect diseases and the presence of pathogens, by identifying the changes in channel impulse responses [37].

Finally, the fourth approach considers ultrasound waves as information carriers [39], where their more favorable propagation in body tissues with high water content shows a potential for enabling communications over distances from several μm to a few cm. Considering the small size of micro-scale nodes, imposing limitations on the employment of standard ultrasound transducers based on the piezoelectric effect and mechanical vibrations, the photoacoustic effect presents itself as a viable option for the excitation of ultrasounds by nano-network devices [40]. The acoustic waves are generated by the thermoelastic mechanism, where light incident on a material surface is absorbed, the rapid heating causing a rapid expansion which in turn generates ultrasonic waves within the material. Therefore, the available nano-lasers can be employed for optoacoustic transmission, while optical resonators can be used for detection at the receiver side [41].

To ensure power supply to both electromagnetic and ultrasound devices, nano-wires made of zinc oxide are proposed for energy harvesting, as they are able to generate electric voltage when bent, e.g., due to a fluid flow in their vicinity. For illustration, about a few thousands of such nano-wires are needed to supply a single nano-machine, each wire being 2 μm long and having 100 nm of diameter [42]. Another proposed approach is powering these machines with remotely generated ultrasounds, which can be converted by them into electrical energy by using piezoelectric nano-elements [43].

While EM nano-communications receive a lot of attention these days, one should note that the required dimensions of the energy sources make this approach suitable for a micro-scale rather than for a nano- one. These systems can be quite easily integrated in networks at the macro-scale, as they share the same communication medium. Consequently, they can act as gateways between macro- and nano-devices. Both EM-based and molecular nano-communication mechanisms are addressed in the IEEE P 1906.1 standard, which has been released in 2015 [44]. The standard defines nano-scale communications, provides a model for ad-hoc nano-communications, and suitable terminology focusing on the nano-communication channel and some higher-layer components, like packets, addressing, routing, localization, and reliability.

As this short survey shows, there is a great variety of available nano- and micro-communication methods, differing in scale, applicable range, and associated delay. While FRET can provide communications between nano-devices in a really short (nano) time scale, its range is limited to about a dozen of nanometers. On the other hand, graphene-based devices working in the THz band are compatible with classical wireless communications, but struggle to operate below the micrometer scale. In between the two, a large number of molecular mechanisms can be exploited. However, these mechanisms are quite distinct from each other, as they use different information carriers, thereby being incompatible with each other, and with FRET and EM-based communication mechanisms. Therefore, the design of appropriate interfaces between nano-, micro- and macro-networks is a critical challenge, and will be a turning point in the further development of nano-communications.

## IV. BODY AREA NETWORKS

Being on the larger and better explored side of the scale, BANs have received considerably more attention over the years than their nano-scale counterpart. Being capable of monitoring their hosts and the surrounding environment, BANs have found their application in a vast range of fields [45], including healthcare, military, sports, and entertainment, among others.

These networks accommodate different types of communications, where the commonly distinguished scenarios include, Fig. 3: among devices inside the human body (in-body), in between devices inside the body and wearable devices on the body (into-body), among wearable devices (on-body), between wearable devices and external access points (off-body), and among wearable devices in different BANs (body-to-body).BANs can use different communication technologies, including wired ones, smart textiles, inductive coupling (ICC), body coupling (BCC), and radio. While some of these technologies are universal and used for all types of BANs, others are limited to certain scenarios or just more convenient, and their employment greatly depends on the application.

While wired communications have certain advantages concerning security and reliability aspects, wires limit user's movement and are prone to material failure, due to constant twisting when the user is dynamic. Therefore, these systems are suitable only for applications with users wearing special suits and involved in low-dynamic activities, e.g., military pilots and Formula 1 drivers [46]. The common drawbacks associated with wires can be overcame by using smart textiles [47], [48], which allow for the integration of power sources, communications, and sensing circuitry within washable clothes [49].

The transmission channel in ICC [50] is established between two magnetic-coupled coils. Since the voltage induced in the receiving coil is inversely proportional to the cube of distance, the coupling exists only for very short distances, i.e., in the order of centimeters. The established channel quality highly depends on the coils' alignment, while being independent of the surrounding tissue. Therefore, ICC is suitable for into-body communications between external devices and near-surface implants, or between nearby implants. An additional advantage of ICC stems from the possibility to use inductive power transfer, where the same coils can provide both the communication link and the power supply for the coupled pair. Since the two impose conflicting requirements on the coil design, the system designer has either to optimize the trade-off in a single-coil design or to use two separately optimized coils [51].

In BCC, the transmitter and receiver couple with the user's body and use it as a transmission medium [52], where one can distinguish between capacitive and galvanic BCC. In the former, the transmitter and receiver couple to the body through capacitive links, created by the electrodes in contact with the skin, while the additional floating electrodes at each side couple with the environment, to provide a return path and close the communication circuit [53]. On the other hand, galvanic BCC exploits ionic properties of the body fluids for signal transmission, with both transmitter/receiver electrodes being in contact with the body [54], [55]. Unlike capacitive BCC, which is limited to on-body communications, galvanic BCC can be used for in- and into-body communications as well. One should note that the field induced by either capacitive or galvanic body coupled transmitters at frequencies above 50 MHz is not quasi-static, and guided wave propagation occurs [56]. The devices are in this case designed to excite strong surface waves traveling along the body [57].

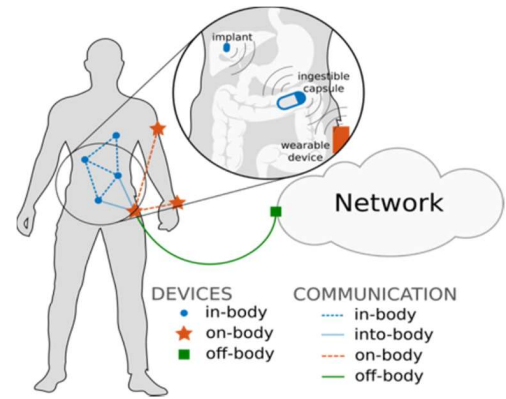

**FIGURE 3.** Communication scenarios in BANs.

Ultrasonic communications use sound waves at frequencies above 20 kHz to transmit information, and their employment in BANs has been proposed only recently [58]. The ultrasounds exhibit a much lower attenuation in body tissues than their EM counterparts [59], making them a more favorable information carrier for in- and into-body communications. Another attractive aspect of this mode of communication is safety, since ultrasounds have been used for medical imaging for a long time, without reported side effects on the human body [58]. Moreover, ultrasound technology has matured, with transceivers being commercially available. One should note that for ultrasound communications, the devices must be in contact with the body, due to poor propagation of ultrasounds trough gasses, therefore, this type of communication is mainly suitable for in- and into-body communications. As in the case of BCC communications, on the one hand, this can be seen as a limitation as contactless off-body communication cannot be accommodated, but on the other hand, it is an advantage in terms of privacy and co-existence, as information is confined to the user's body.

While ultrasound communications provide data rates only up to 100 kbit/s, the employment of higher order modulation schemes, such as 64 QAM [60], or spatial-multiplexing MIMO [61], has been proposed to achieve data rates higher than 10 Mbit/s. However, the transceiver structure complexity and energy consumption associated with these schemes could be beyond the reach for implanted devices, due to their small-size and constrained energy source.

In addition to sound, light can be also used as an information carrier in BANs. Besides ton-body communications [62], optical wireless transmission can be also employed to establish links between implants and on-body nodes [63], where data rates of up to 50 Mbit/s have been demonstrated through 4 mm thick tissues [64]. However, transdermal optical communications can be established with devices implanted several centimeters into the body tissue [65], by using wavelengths within the spectral window between 700 and 950 nm, characterized by low absorption [66]. With the power constraints of implanted devices being a critical point, the extension of the longevity of implants can be achieved by exploring retroreflective transdermal communication for the transmission of data from an implant to an outside device [66]. In this case, the light source is provided by a less constrained on-body device and the implant modulates the reflected light by a micro-electrical mechanical system (MEMS) device, i.e., a miniature modulating retroreflector (MMR). In addition to this unique feature of transdermal optical communications, the immunity to EM interference and inherent security due to limited propagation range are some of the advantages of this technology.

Although the aforementioned communication systems are associated with certain advantages, radio is by far the most considered option for BANs, due to the convenience of wireless transmission and the availability of a number of mature technologies, namely, Bluetooth [67], Bluetooth Low Energy (LE) [68], ZigBee [69], and ultra-wideband (UWB) [70]. These technologies are also adopted by the physical layer specifications in the existing standard for BANs, i.e., IEEE 802.15.6 [71]. Table 2 gives a summary of the communication technologies, indicating the associated frequency bands, ranges, and throughputs. Bluetooth was initially designed for the exchange of audio and data between personal devices, but its range and available data rates make it suitable for off-body transmission of aggregated BAN data. Bluetooth LE was later developed as its energy efficient, low latency, and low-cost alternative, being compatible with Bluetooth devices and suitable for communications with sensor devices. ZigBee offers another low-energy and low-cost solution for BAN communications, at the expense of a lower available data rate. By allowing for multi-hop transmission to circumvent propagation path obstructions, ZigBee can extend nodes' coverage and improve the communication reliability.

However, UWB is probably the most popular radio technology for BANs, due to its advantages associated with the large bandwidth and low power density (below -41.3 dBm/MHz). The latter makes UWB suitable for operation in environments sensitive to electromagnetic radiation, and in the close vicinity of the human body, as it yields low exposure levels. Moreover, the large corresponding penetration depths make UWB a popular choice for in- and into-body communications.

One should also note that millimeter-waves BAN communications have been considered recently, due to the high capacity and low inter-BAN interference [72]. The propagation characteristics of millimeter waves are both attractive and challenging for BANs. On the one hand, the high propagation losses are favorable for BAN coexistence, secrecy and security reasons, as signals remain confined within the area of the user's close proximity. On the other hand, these losses pose a great challenge on preserving the information-bearing signal above the level required for satisfying communication quality. Millimeter waves BANs are still a niche research area, but the enabling technologies, such as 60 GHz WLAN (WiGig) [73], already exist.

TABLE 2. Summary of main communication technologies for BANs.

| Physical link | System/Mechanism | Freq. [GHz] | Range [m] | Data rate [Mbit/s] |
|---|---|---|---|---|
| Wired | Smart-textile | - | - | 0.25 |
| Wireless | Optical comm. | $10^5$ | 0.05 | 50 |
| | ICC | < 0.02 | 0.2 | 2.5 |
| | Ultrasound | < 0.3 | 0.5 | 0.1 |
| | Capacitive BCC | < 0.2 | 1 | 10 |
| | Galvanic BCC | < 0.05 | 1 | 0.064 |
| | Bluetooth LE | 2.5 | 10 | 1 |
| | WiGig | 60 | 10 | 7 000 |
| | UWB | 3-10 | 50 | 480 |
| | Bluetooth | 2.5 | 100 | 3 |
| | WiFi | 2.5, 5.8 | 100 | 100 |
| | ZigBee | 2.5 | 200 | 0.25 |

While BANs show a great potential, these systems are faced with a number of challenges, including energy efficiency, reliability, user safety, data security, simple use and comfort. These matters have to be addressed at all design levels, i.e., from physical transmission, medium access, and routing protocols, to top-level software design. Some of the most important challenges are associated with the communication channel, and the peculiarities of body-centric radio propagation, where body-shadowing and users' motion have a significant influence on channel quality [74], [75]. Moreover, the human body strongly affects the radiation characteristics of antennas operating in its close vicinity, which imposes great challenges on antenna design, especially in the case of in- and into-body communications, where antennas are immersed in highly dispersive human tissues [76], [77].

Despite the challenges, BANs have already demonstrated great potential for personal systems. The ability to seamlessly monitor the host and communicate with the surroundings has already made BANs a part of medical treatments in modern hospitals, and a helping hand in state-of-the-art training for athletes. However, their true potential extends far beyond the current systems, and could be achieved through a collaboration between nano-networks and BANs, allowing for information to be carried from a tissue inside the human body to the outside world. In such a scenario, a nano-network will transfer information to an implant, where a BAN would carry it to an off-body access point, from where a large-scale network would be in charge of delivering it to a remote location. The implications are truly mind-boggling, when one considers applications in healthcare or bio-metrics, as well as those that are beyond one's imagination today. The key enablers of such truly personal communications are the interfaces between devices and mechanisms at different scales. An overview of some mechanisms that can be exploited for this purpose is given in the next section.

TABLE 3. Potentials and limitations of interfacing mechanisms

| Interfacing mechanism | Source | Destination | Potentials | Challenges |
|---|---|---|---|---|
| Channelrhodopsins | FRET networks | Neurons / EM micro-devices / BANs | Light-to-voltage converters | Limited throughput, below 1 bit/s for a single ChR, 200 bit/s in neuron |
| Photodetectors | FRET networks | BANs | Very high potential throughput of Mbit/s | Low photodetection efficiency < 50% |
| BRET | Molecular comm. | FRET networks | Local energy source | Throughput of few bits/s |
| ATP | Molecular comm. | FRET networks / neurons | Extremely ubiquitous molecules | Difficult to be controlled |
| EM communication | EM micro-devices | BANs | Two-way communication Millimeter ranges | THz or higher frequencies required |
| SPR sensors | BANs | EM micro-devices | High throughput | Specific incidence angle of the hitting EM wave required |
| BAN-controlled light diodes | BANs | FRET networks | Very high potential throughput of Mbit/s | Line-of-Sight conditions required |

## V. NANO-MICRO-MACRO INTERFACES

While numerous communication mechanisms in both nano- and micro-scales have been proposed and thoroughly investigated, they are still quite disconnected from each other. It is mainly because they use different information carriers, e.g., FRET/photons, THz waves, diffusing molecules, and DNA strands. Consequently, it is challenging to transfer data from one nano-network type to another, or from nano- and micro-scales to the macro-world of conventional wireless devices. Still, some interfaces can be established by exploiting the physical properties of specific molecules or materials, which are presented in the following subsections and summarized in Table 3. In Fig. 4, a visual understanding of mechanisms mentioned in subsections A-D is also given, while in Fig. 5 two cases of micro-macro EM communications are illustrated, as discussed in subsection E.

### A. LIGHT-STIMULATED CHANNELRHODOPSINS

Channelrhodopsins are small protein molecules, with a size of about 5 nm, naturally occurring in some green algae organisms, able to open light-induced channels for ions. Recently, they became highly investigated [78], [79], inspiring quite a new discipline called optogenetics, which is about controlling living tissues with light. Channelrhodopsins, when treated with light, have an ability of opening a pore (channel) where positive ions, e.g., present in blood, can flow through. Such a channel remains open by at least 10 ms, which is enough for the flowing ions to change the electrical potential at the other side of the ion channel. Furthermore, channelrhodopsins can be activated by FRET as well [80]. From the nano-communications viewpoint, channelrhodopsins can be then understood as light-to-voltage converters. As the voltage can be later measured by an EM-based micro-device or a BAN, channelrhodopsins constitute an interface between these networks, being able to transfer FRET signals to electrical ones. The full cycle of a channelrhodopsin, from closed to open and then closed and able to accept stimuli, is as long as 5 s [81], thus, the throughput of data transferred via a single channelrhodopsin remains far below 1 bit/s. A solution can be a large layer of channelrhodopsin molecules working in parallel [80], which has been shown to have a throughput of about 50 bit/s, achieved with a bit error ratio kept below $10^{-3}$.

Channelrhodopsins can be also embedded into neurons, replacing receptors at the neuron membrane, which has already been proved experimentally [82], [83]. Like neuron receptors are able to receive stimuli from tens of thousands of other neurons via neurotransmitters, similarly channelrhodopsins signals can be received from many sources. Such a neuron is then stimulated by light or FRET. It has been shown that channelrhodopsin-controlled neurons are able to produce up to 200 impulses per second [84], which means about 200 bit/s for the respective channel throughput.

### B. LUMINESCENCE BY BRET

BRET stands for bioluminescence resonance energy transfer and is a process similar to FRET, as explained in Section III, although the donor energy does not come from external excitation, but from a chemical reaction. In this reaction, a luciferase molecule, like Vluc, Rluc, Fluc or NanoLuc, is oxidized in the presence of a specific substrate molecule, a luciferin. The reaction produces energy that is emitted as a photon or can be caught by a donor molecule, if situated nearby. From the communication viewpoint, BRET can initiate a single or multi-hop FRET transmission without any external source of donor excitation. As the mentioned reaction is strictly dependent on the suitable amount of luciferin substrates, it can be controlled by molecular diffusion, for example. Thus, BRET has two crucial advantages: (a) local origin via chemical reaction, and (b) control with a molecular communication mechanism, i.e., diffusion. Consequently, BRET can intermediate between molecular communication mechanisms and FRET transmissions.

BRET has its limitations related to the physical properties of luciferases and to the chemical reaction characteristics. First, the reaction turnover rate is quite low: for NanoLuc luciferases it is 0.5 s, while for RLuc it is even 5 s [85]. For comparison, a FRET donor molecule is ready for energy

IEEE Accessabsorption just after a previous emission, which usually occurs in a few nanoseconds. In effect, a single luciferase can pass only around 1 bit/s, comparing with over 10 Mbit/s for a typical FRET transfer. Second, similarly to FRET, communication distances for BRET are below 10 nm [80].

### C. THE ROLE OF ATP

ATP (adenosine triphosphate) is the most common energy currency for living cells. It is extremely ubiquitous: despite that a single ATP is about 1 nm in size, each human body produces more of these molecules during the day than the whole-body weight. An ATP molecule consists of an adenosine and three highly energetic phosphate groups. Living organisms usually spend ATP energy via its hydrolysis, removing one phosphate group and thus creating an ADP (adenosine diphosphate) molecule; in this process 30.6 kJ is released per mole of ATP.

While being so common energy carriers, ATP molecules are also proved to be information carriers between living cells. ATP can act similarly like neurotransmitters, which, in biology, is called purinergic signaling [86], [87]. Specific receptors on neural cells are already known, namely P2X and P2Y, matching strictly ATP molecules. The former opens an ion channel on a cell membrane, like a typical neuron receptor, while the latter triggers releasing calcium stores inside the cell. Thus, ATP molecules can carry information from a cell to cell via diffusion, like neurotransmitters.

ATP can initiate communication not only to nerve cells, but also to BRET/FRET networks. ATP is required for chemical reactions starting BRET with, e.g., Fluc (firefly luciferase) molecules, so it can be used to control these reactions [88]. Also, ATP is the energy source for the movement of kinesins and dyneins. These molecular motors advance 8 nm for each ATP molecule spent [89], [90], so this way of transporting information can be clearly controlled via careful ATP delivery.

As ATP is so important for nano-communications, its production should be controlled, which, fortunately, is feasible. ATP can be released by neural cells, like neurotransmitters, but, as previously mentioned, ATP is also produced from ADP by all living organisms. In plants and some bacteria, it is done during the light part of the photosynthesis, with a reaction called photophosphorylation, using the energy of light. In non-photosynthetic organisms, e.g., humans and animals, ATP is produced in catabolic processes of so-called cellular respiration via oxidation of carbon-containing structures, like fatty acids or carbohydrates, e.g., from a single glucose $C_6H_{12}O_6$ molecule, living cells produce up to 38 ATP molecules [91].

Summing up, ATP molecules can intermediate between different types of communications. ATP molecules can be provided directly, or their production can be stimulated with light (light-induced) in photosynthetic structures or by delivering carbon-containing compounds to non-photosynthetic cells. ATP initiates BRET communications, gives fuel for movement of molecular motors or fires actions of cells having P2X/P2Y receptors, neuronal actions in particular.

### D. PHOTODETECTORS AND SPR

Recently, an innovative system has been presented, where genetically modified Escherichia coli bacteria integrated with a wireless endoscopy capsule acquired information regarding internal bleeding inside gastrointestinal tract [92]. The bacteria signaled the information via luminescence to the photodetectors designed for this purpose [93]. The system has been successfully tested with in vivo experiments and is an example of an efficient communication interface, via light, between a molecular system and macro-scale devices.

The bacteria mentioned above send a very simple information, just a confirmation that a bleeding is located, without any bit transmission. Photodetectors technology has progressed very rapidly in the last twenty years, and currently these devices, commercially available, are able to intercept/collect much more subtle signals; in particular, the so called single-photon detectors (SPDs) look very promising, as they are able to count and time-stamp incoming photons. These devices are therefore able to receive bit transmissions from FRET-based nano-networks, as FRET-receivers emit received signals as photons.

Among many families of SPDs, not all of them are suitable for creating networking interfaces. Real photodetectors are characterized by intrinsic limitations, resulting from the specific physical mechanism of photodetection. Popular solutions, such as superconducting nano-wire SPDs or transition-edge sensors, are working in cryogenic temperatures only, in order to reduce background noise. On the other hand, much more suitable are single-photon avalanche diodes (SPADs), able to perform at room temperatures. SPAD photodetectors are commonly used to measure FRET efficiency in biophotonic applications [94]. The size of the active area of a SPAD is usually a few dozens of micrometers, but if larger dimensions are required, then SPAD arrays are commonly created. In a SPAD, an absorbed photon creates an electron-hole pair that is, in turn, multiplied in the avalanche process. This process must be further stopped in order to make the SPAD sensitive for the next photons, which usually takes 10 to 100 ns [94], which means that photons can be accepted with a rate of the order of 10 Mbit/s. While the evolution of SPAD technology is progressing well, the main challenge is their photodetection efficiency, which rarely exceeds 50% and is not stable in the whole visible and infrared spectrum [94]. In consequence, probably, at least a few absorbed photons are required per each bit of data (repetition coding), which slows down the communication throughput of this technology. Recently, a single-photon detector based on cadmium sulfide nano-wire, but operating in room temperatures, has been also presented [95] and can be treated as an alternative for SPADs. Summing up, single-photon detectors, especially SPADs, are a developing technology that can be seriously considered as a solution for interfacing FRET-based and larger-scale networks.

When discussing optical interfacing solutions, Surface Plasmon Resonance (SPR) should be also mentioned. The SPR phenomenon occurs when an optical/infrared/THz wave hits a metallic or graphene surface. If specific conditions regarding the incidence angle and the wave frequency are met, a rapid oscillation of electron density is created, which propagates along the surface of the metal/graphene. This oscillation is called a plasmon and it can travel about 10 µm [28] in graphene or up to 60 µm in silver/gold [96]. Classical prism-based SPR sensors are quite large, but solutions based on optical fibers are much more compact with sizes of about 20 µm [97]. As the conditions for plasmon creation depend also on the environment's parameters, SPR sensors are commonly used to study biological structures. Yet, the SPR phenomenon can be considered as a method for converting optical or infrared signals into plasmons that in turn can be detected by micro-scale graphene devices. SPR sensors can achieve very good response times of about 1 µs [98], thus the throughput of such an interface might be in the order of 1 Mbit/s. However, a limiting factor is the specific incidence angle of the wave hitting the graphene surface: it requires a precise geometric configuration of the whole interface system, which probably limits its application to the case of a macro-BAN node transmitting to a static graphene micro-device.

### E. MICRO-MACRO EM COMMUNICATIONS

Communications between micro-scale nodes and an external BAN can be also established by employing EM waves at THz or optical bands as information carriers. The extremely high path loss in the presence of multiple tissues in between transmitter and receiver make a direct communication between a nano-node in the blood stream and an external BAN device particularly challenging [99]. However, communication with external BAN devices could be achieved by introducing mediator devices, assisting the information delivery from micro-scale devices circulating through the blood stream to an on-body macro-scale device placed on the skin.

These devices could be micro-scale nodes implanted in between blood vessels and the skin, employing THz, optical or ultrasound waves to communicate with both sides [100], [101]. Both the micro-scale devices carrying information and the implanted one can be powered by energy delivered by ultrasound waves emitted by the on-body BAN node [101]. The information exchange can then take place while the circulating nano-node is within the portion of the blood vessel in which it can harvest the ultrasound waves' energy.

However, the location of these mediating devices must be chosen carefully, so that favorable propagation conditions and a high chance of establishing contact with nano-nodes circulating through the body are ensured. The former requirement means that the body parts with thin tissue layers between blood vessels and the skin are preferable, while the latter suggests that a micro-scale gateway should be placed next to veins and arteries that see most of the blood volume passing through within a short period of time. Moreover, except possibly for critical applications with users in hospital beds, the mediator node location is also constrained by aesthetics and the practicality of on-body device placement, considering daily life activities.

For example, the internal jugular veins in the neck provide a high level of interaction with nano-nodes circulating through the body, as they host 14% of the total blood-flow in the body [102], but the attachment of an external node on the neck is inconvenient for regular daily activities. On the other hand, the veins in the wrist are attractive for their proximity to the surface, i.e., thin tissue and superior accessibility, since the external collector device could be integrated within a watch or a bracelet, but only 1% of the total blood flows towards the heart through these veins [103].

Alternatively, the mediating devices could be in-body BAN nodes that can use light, THz waves or ultrasound to communicate with micro-scale devices, and microwaves with devices on the body surface, or even with nearby off-body devices. Since these devices are larger and more capable than their micro-scale counterparts, their position is less constrained by the distance from the on-body node and can be better optimized in terms of proximity to blood vessels with a high probability of interaction with circulating nano-nodes. Propagation through body tissues at microwaves is much less restricted than in the THz band, but the high dispersion losses due to conductivity still limit communication to distances of the order of 1 cm of body tissue [104]. However, at microwaves, communications can be used to deliver information from implants directly to a device in the user's proximity, such as a medical instrument next to the hospital bed [105], [106]. UWB communications seem to be the most attractive option for into-body communications, due to the high tissue penetration depths [76], [77], but other technologies discussed in Section IV could be employed.

Instead of microwaves, ultrasonic communications can be also used for into-body communications with BAN implants in this scenario [58], or even directly with micro-scale devices. Ultrasounds are characterized by lower absorption losses, and their proved safety through the long usage in medical imaging is another attractive property [59]. Similarly, transdermal optical communications can be used to establish links with in-body BAN or micro-scale devices, however, with more strict limitations than with ultrasound carriers. The direct optical link between micro-scale and on-body devices is particularly challenging, and unlikely to be established with optical EM communications in the outward direction.

### F. SUMMARY AND RESEARCH CHALLENGES

Summing up the solutions discussed above, one can say that two trends are dominating. First, traditionally in communications, wireless solutions are extensively popular, as the wireless medium is now ubiquitous, having in mind all tiny sensors and IoT devices. The second group of solutions is based on light, which, on the one hand, is still an EM wave, but, on the other hand, an energy carrier commonly used in biological systems. Except for the cases already described, i.e., channelrhodopsins, luminescence and photodetectors, optical signals, e.g., emitted by BAN-controlled diodes, can easily intermediate between macro- and nano-networks. Due to the scale difference, the signals coming from macro-devices are of a broadcast nature for nano-networks, but can be filtered by properly designed nano-devices. For example, one can consider using an optical source, such as a laser, to transfer signals into FRET-based nano-networks, where the nano-transceivers can be fluorophores, such as Alexa, DyLight or Atto selective dyes, i.e., commercially available bio-engineered molecules. There is a wide variety of these dyes, characterized by different absorption spectra in the visible light range. from close to ultraviolet (380 to 400 nm), up to nearly infrared (700 to 750 nm).

They can be used for the frequency-selective reception of optical signals, similarly as in standard wavelength division multiplexing (WDM) receivers. Optical signals can be also received by molecular communication systems, where the properties of bacteria, such as Escherichia coli, can export protons (H+) and change the pH value of its environment in response to optical stimulation [107].

The approaches presented in the above subsections prove that despite nano-, micro- and macro-communication mechanisms are quite distinct from each other, some lessons from physics, optogenetics, biotechnology, and wireless communications can be learned and then interfacing techniques are at one's fingertips. However, one should be aware of the low level of their maturity: the physical solutions and biological components are ready, but still, there is much engineering work to be done to have the whole inter-scale communications systems operational. Except for the research challenges specific for the chosen interfacing techniques, there are at least three general challenges pivotally important for the whole subject:

a) Interdisciplinary cooperation - Understanding the mechanisms and phenomena that are the basis for inter-scale communications requires knowledge from many scientific areas. While information and communication technologies are in the center of this research, physics, biotechnology, medicine, and optogenetics are crucial for making progress on interfacing solutions. Cooperation among these disciplines is strenuous, as different research communities describe phenomena with a distinct language and are used to focus their work on different goals. However, such a cooperation, much needed, will result in a substantial progress and a holistic expertise on this whole subject.

b) Experimental effort - Most of the current research in nano-communications is performed via a theoretical approach or computer simulations, sometimes not using realistic assumptions. This subject requires more experimental work in many areas, e.g., biological samples or physical nano-devices. While in BANs there are relatively more experimental studies, they do not concern interfaces with nano- and micro-scale networks.

c) Control over device positions and mobility - Many of the discussed interfacing techniques require a precise

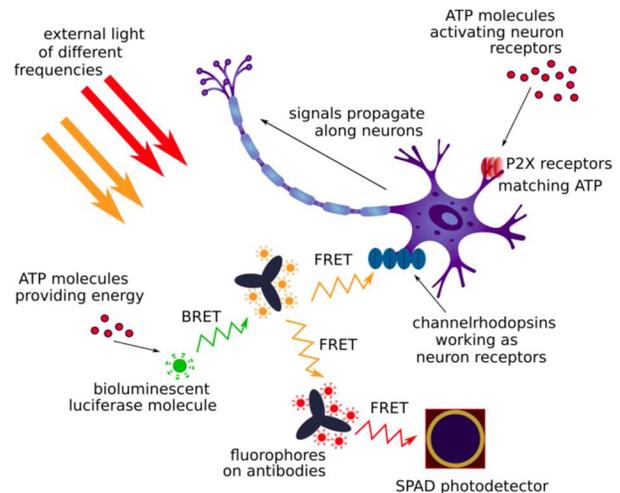

**FIGURE 4.** Communication interfaces for FRET and molecular networks (the device scale is not preserved).

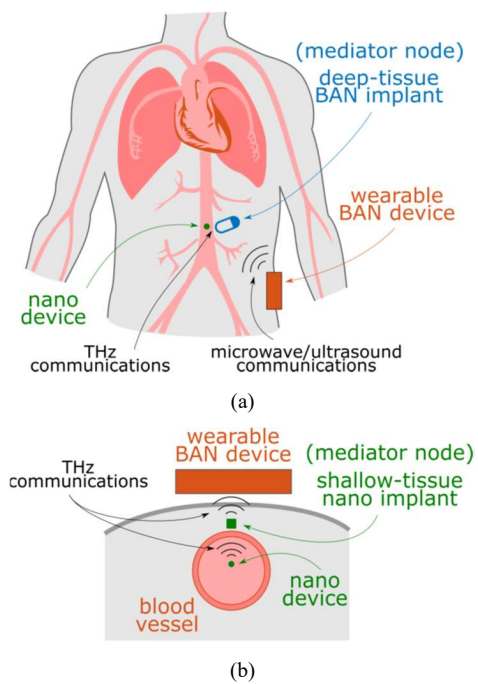

**FIGURE 5.** Micro-macro EM communications with BAN and nano device mediator nodes. (a) BAN implant as a mediator node. (b) Nano-device as a mediator node.

control over the relative positions of the communicating nodes, e.g., FRET networks must be exactly positioned so that their signals are received by channelrhodopsins or ATP molecules should be carefully delivered to initiate BRET. Such bio- or nano-engineering control over molecules and nodes in their environment is one of the most important barriers on the way to the development of interfacing solutions.

In general, communications systems and networks are described by a number of parameters and entities, which depend on the nature of the communication and sometimes even on the specific application. The description of transmitters and receivers, and the characterization of the channel in between them are basic requirements for any system, while architectures, routing algorithms and protocols form some of the bases for networks, and this paper provides some information in this direction. While interfaces have been addressed, as well as basic transmission and reception, together with some coverage ranges and data rate capacity, other matters are still open for research, such as models for channels, interference, connections capacity, routing, addressing, reliability, and efficiency, among many others.

## VI. CONCLUSIONS

Nano-communications between groups of molecules, cells, or EM-based micro-devices, are one of the hottest research areas in communications in recent years. The progress in this discipline is strenuous, since it requires cooperation between IT and medical experts, but the potential applications are countless, extending from healthcare, such as remote diagnostics, surgery and localized drug delivery, to sports, entertainment and military, while encompassing Human-Internet integration. The latter is a vision of truly personal communications, where the far edges of the communication network are deeply inside people's bodies, allowing for delivering the signals directly from human nerve cells to the outside network.

This paper presents a brief survey of important nano-communication mechanisms, including molecular mechanisms, Förster resonance energy transfer phenomenon and EM micro-devices. It also provides an overview of the BAN communication technologies, including smart-textiles, inductive- and body-coupling mechanisms, and a number of available suitable radio technologies (e.g., UWB, Bluetooth, ZigBee, and WiFi). However, the main contribution of this paper is in the summary of potential interfacing mechanisms between nano-networks and BANs, which will serve as mediators to the outside world. These interfaces are the critical aspect of the envisioned truly personal communication systems and require dedicated research efforts to fully realize their potentials.

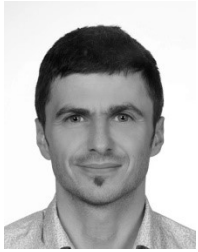

**PAWEL KULAKOWSKI** received his Ph.D. in telecommunications from the AGH University of Science and Technology in Krakow, Poland, in 2007, and currently he is working there as an assistant professor. He spent about 2 years in total as a post-doc or a visiting professor at Technical University of Cartagena, University of Girona, University of Castilla-La Mancha and University of Seville. He was involved in research projects, especially European COST Actions: COST2100, IC1004 and CA15104 IRACON, focusing on topics of wireless sensor networks, indoor localization and wireless communications in general. His current research interests include molecular communications and nano-networks. He was recognized with several scientific distinctions, including 3 awards for his conference papers and a governmental scholarship for young outstanding researchers.

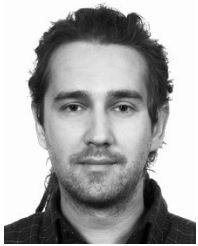

**KENAN TURBIC** (S'14–M'19) has received his MSc degree from the University of Sarajevo in 2011, and a PhD degree (Hons.) in Electrical and Computer Engineering from IST, University of Lisbon, in 2019. He is currently a postdoctoral researcher at the INESC-ID research institute, Lisbon, Portugal. His main research interests are wireless channel modelling, with a particular interest in Body Area Networks. He is actively participating in the COST Action CA15104 (IRACON), to which he has contributed with several technical documents and is serving as one of the Section editors for the final report book.

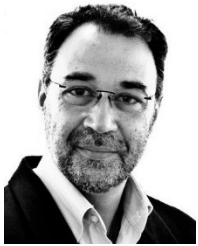

**LUIS M. CORREIA** (S'85-M'91–SM'03) was born in Portugal, in 1958. He received the Ph.D. in Electrical and Computer Engineering from IST (University of Lisbon) in 1991, where he is currently a Professor in Telecommunications, with his work focused on Wireless & Mobile Communications in the areas of propagation, channel characterisation, radio networks, traffic, and applications, with the research activities developed in the INESC-ID institute. He has acted as a consultant for the Portuguese telecommunications operators and regulator, besides other public and private entities, and has been in the Board of Directors of a telecommunications company. Besides being responsible for research projects at the national level, he has participated in 32 projects within European frameworks, having coordinated 6 and taken leadership responsibilities at various levels in many others. He has supervised more than 200 M.Sc./Ph.D. students, having edited 6 books, contribute to European strategic documents, and authored more than 500 papers in international and national journals and conferences, for which served also as a reviewer, editor and board member. Internationally, he was part of 36 Ph.D. juries, and 66 research projects and institutions evaluation committees for funding agencies in 12 countries, and the European COST and Commission. He has been the Chairman of Conference, of the Technical Programme Committee and of the Steering Committee of various major conferences, besides other several duties. He was a National Delegate to the COST Domain Committee on ICT. He was active in the European Net!Works platform, by being an elected member of its Expert Advisory Group and of its Steering Board, and the Chairman of its Working Group on Applications, and was also elected to the European 5G PPP Association. He has launched and served as Chairman of the IEEE Communications Society Portugal Chapter.